\documentclass[twocolumn,english,conference]{IEEEtran}
\usepackage[T1]{fontenc}
\usepackage[latin9]{inputenc}
\usepackage{units}
\usepackage{mathtools}
\usepackage{algorithm2e}
\usepackage{amsmath}
\usepackage{amsthm}
\usepackage{amssymb}
\usepackage{graphicx}

\makeatletter

\providecommand{\tabularnewline}{\\}

\theoremstyle{plain}
\newtheorem{thm}{\protect\theoremname}
\theoremstyle{remark}
\newtheorem{rem}[thm]{\protect\remarkname}
\theoremstyle{plain}
\newtheorem{lem}[thm]{\protect\lemmaname}
\theoremstyle{plain}
\newtheorem{prop}[thm]{\protect\propositionname}

\usepackage[hang,flushmargin]{footmisc}
\usepackage{babel}
\usepackage[font={small}]{caption}
\LinesNumbered
\SetKwRepeat{Do}{do}{while}%

\providecommand{\lemmaname}{Lemma}
\providecommand{\propositionname}{Proposition}
\providecommand{\remarkname}{Remark}
\providecommand{\theoremname}{Theorem}

\usepackage{microtype}

\providecommand{\lemmaname}{Lemma}
\providecommand{\propositionname}{Proposition}
\providecommand{\remarkname}{Remark}
\providecommand{\theoremname}{Theorem}

\newcommand\blfootnote[1]{%
  \begingroup
  \renewcommand\thefootnote{}\footnote{#1}%
  \addtocounter{footnote}{-1}%
  \endgroup
}

\@ifundefined{showcaptionsetup}{}{%
 \PassOptionsToPackage{caption=false}{subfig}}
\usepackage{subfig}
\makeatother

\usepackage{babel}
\providecommand{\lemmaname}{Lemma}
\providecommand{\propositionname}{Proposition}
\providecommand{\remarkname}{Remark}
\providecommand{\theoremname}{Theorem}

\begin{document}

\title{A Novel Framework for Modeling and Mitigating Distributed Link Flooding
Attacks}

\author{Christos Liaskos$^{1}$, Vasileios Kotronis$^{2}$ and Xenofontas
Dimitropoulos$^{1}$\\
 {\small{}$^{1}$FORTH, Greece\quad{}\quad{}$^{2}$ETH Zurich, Switzerland}\\
{\small{} Emails:~\{cliaskos,~fontas\}@ics.forth.gr,~vkotroni@tik.ee.ethz.ch}}
\maketitle
\begin{abstract}
Distributed link-flooding attacks constitute a new class of attacks
with the potential to segment large areas of the Internet. Their distributed
nature makes detection and mitigation very hard. This work proposes
a novel framework for the analytical modeling and optimal mitigation
of such attacks. The detection is modeled as a problem of relational
algebra, representing the association of potential attackers (bots)
to potential targets. The analysis seeks to optimally dissolve all
but the malevolent associations. The framework is implemented at the
level of online Traffic Engineering (TE), which is naturally triggered
on link-flooding events. The key idea is to continuously re-route
traffic in a manner that makes persistent participation to link-flooding
events highly improbable for any benign source. Thus, bots are forced
to adopt a suspicious behavior to remain effective, revealing their
presence. The load-balancing objective of TE is not affected at all.
Extensive simulations on various topologies validate our analytical
findings.\blfootnote{This work was funded by the European Research Council via Grant Agreement no. 338402, project ''NetVolution: Evolving Internet Routing''.}\end{abstract}

\begin{IEEEkeywords}
DDoS, link-flooding, analysis.
\end{IEEEkeywords}

\IEEEpeerreviewmaketitle{}

\section{Introduction\label{sec:Introduction}}

DDoS attacks are well-known within the Internet community. For example,
the attack against Spamhaus in 2013 was a powerful attack that inflicted
more than $300$ Gbit/s of malicious traffic upon the intended target~\cite{spamhaus}.
However, researchers have recently shed light on new types of link-flooding
attacks that can operate at Internet scales and are extremely difficult
to detect and mitigate~\cite{crossfire,Coremelt}. The objective
of these attacks is to deplete the bandwidth of certain network links,
disconnecting entire domains---even countries---from the Internet.
In particular, the Crossfire~\cite{crossfire} attack is a stealthy
and effective DDoS link-flooding attack that uses bots with non-spoofed
IP addresses to send traffic to publicly accessible servers. While
packets stemming from these attack sources are seemingly legitimate,
their cumulative volume harms the intended victim in an indirect way
by flooding links and cutting off connectivity towards its location.
Each bot-to-server flow usually has very low bandwidth, and is thus
very hard to detect and filter.

On the defender's side, Traffic Engineering (TE) is the network process
that reacts to link-flooding events, regardless of their cause~\cite{Pioro.2004}.
The TE module, hosting this process, is thus a natural point to incorporate
attack detection and mitigation mechanisms. A TE process has two phases:
i) the optimal load calculation for each network path, and ii) the
mapping of specific traffic flows to paths in a manner upholding their
calculated optimal load \cite{Balon.2006b}. The first phase represents
the first TE priority, which is to ensure that the network load is
balanced, i.e., fairly distributed over all available paths. The flow
mapping phase has received limited focus in existing TE solutions
and is even random for all but elephant flows~\cite{Curtis.2011b}.

In the present paper the TE process is optimized for attack detection,
without altering its load-balancing objective. The \emph{methodology}
consists in optimizing the flow mapping phase of TE. The new optimal
mapping ensures that a benign flow will have the lowest probability
of contributing to a future attack by chance. Therefore, bots are
forced to behave improbably over time in order to remain effective.
Internally, a novel analytical framework based on relational algebra
is employed to relate bots to susceptible targets. In this aspect,
the outlined TE flow mapping maximizes the support of bot-to-target
relations, accentuating their presence over time.

To the best of our knowledge, this paper \emph{contributes} the first
analytical model targeting Crossfire-like link-flooding attacks, offering
novel insights in this attack type. In addition, the presented analysis
and algorithm have general applicability to multigraphs, multipath
routing and generic bot behavior. Multiple time-varying and mixed
malicious/benign connections per single bot are allowed. Furthermore,
the analysis offers insights in the topological attributes that affect
the attack and its mitigation. Moreover, the integration with existing
TE modules is seamless. Finally, given that the analysis is built
upon relational algebra principles, the proposed scheme lends itself
to a straightforward implementation based on well-known, mature and
scalable databases (e.g., SQL \cite{SQLbook}).

The remainder of this paper is organized as follows. A description
of the studied attack is given in Section \ref{sec:The-Attack-Model}.
The proposed analytical framework is described in Section \ref{sec:Analysis},
while its algorithmic formulation is given in Section \ref{sec:Algorithmic-formulation}.
The simulation results follow in Section \ref{sec:Simulations}, while
the related work is presented in Section \ref{sec:Related-Work}.
Finally, we conclude in Section~\ref{sec:Conclusions}.

\section{Attack Model: Reactive Crossfire\label{sec:The-Attack-Model}}

\begin{figure}[t]
\begin{centering}
\includegraphics[width=0.9\columnwidth]{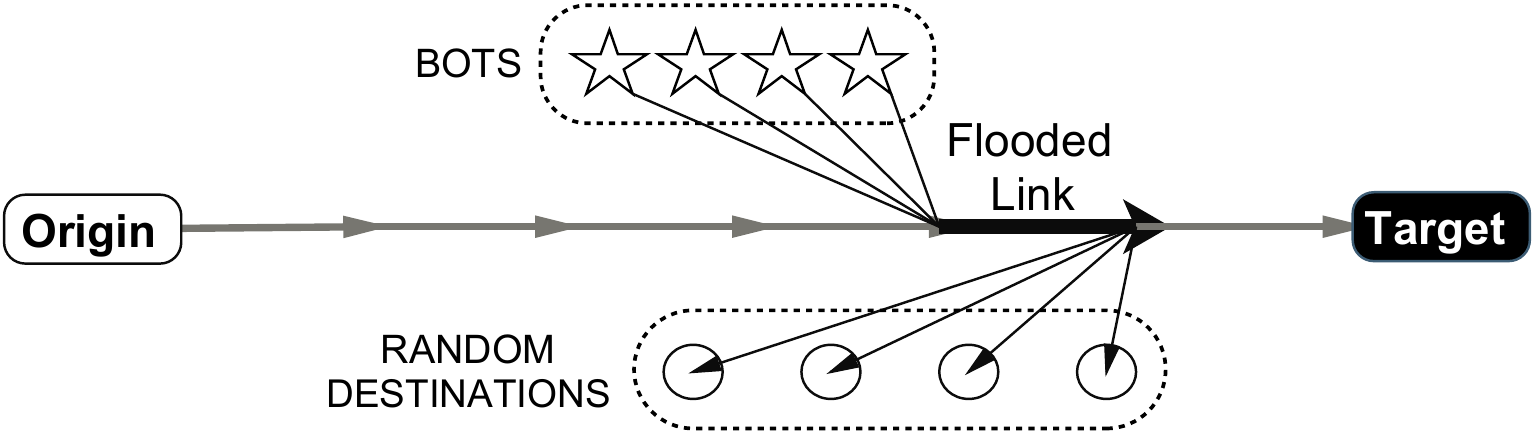}
\par\end{centering}

\protect\protect\caption{{\footnotesize{}\label{fig:crossfireOverfiew}Overview of the Crossfire
attack concept.}}
\end{figure}

The attack that we study as a use case is a reactive version of the
Crossfire attack. In the classic Crossfire attack, the attacker has
a swarm of bots (or botnet) at his disposal, and seeks to attack a
certain area, called the \emph{target area}. The goal is to cut off
Internet connectivity to this area. To achieve this, he assigns his
bots to send legitimate, low-rate traffic flows towards certain public
servers, the \emph{decoy servers}. These servers are reached over
the same \emph{target links} that connect the target area to the Internet.
Thus the bots send traffic along paths that lead to both the decoys
and the target, cumulatively flooding the shared target links. Traffic
is sent only to the decoys, so that the target cannot directly observe
the flood. The knowledge base of the attacker is mainly the \emph{link-map},
i.e., the map of the links of the victim network. The most loaded
links along the paths from the bots to the servers and from the bots
to the target are flooded. The concept is illustrated in Fig. %
\mbox{%
\ref{fig:crossfireOverfiew}%
}; the decoys can be random destinations that are on the appropriate
paths to the target. As an extension, we also assume that the attacker
monitors the network routes and reacts to any changes in the link-map,
including shifts of load and routing changes. These changes may be
possible defender's reactions; the attacker thus reiterates this process
to harm the target anew.

While the attacker seeks to flood the target links, the defender aims
at alleviating the load from the flooded connections and at finding
and blocking any malicious traffic sources. While the first objective
is the same as the one employed by most modern TE modules (i.e., balancing
the load), the second one is much harder to implement. In our framework,
we thus assume the following features for the defender's process.
First, he monitors the network load and reacts to link-flooding events.
Flood detection can be based on approaches such as the one from Xue
et al.~\cite{xue2014towards}. After the flood is known, the defender
balances the load by re-routing traffic destined to different destinations,
without though knowing the attacker's classification (target, decoys,
benign servers). Moreover, he records sources that are consistently
present in link-flooding events, even after re-routing. Sources that
change their destination selection to adapt to re-routing are particularly
suspicious; that means that re-routing has diverted their initial
load away from the target link(s), while they want to return and inflict
damage. The idea is that after such an interaction cycle between the
attacker and the defender, attack sources may become more identifiable
by exhibiting a behavior that is highly improbable for benign sources.
For example, benign (e.g., flash-crowd) load would not re-adjust to
routing changes, but it would use the same popular destination(s)
as before.

The defender must gradually collect evidence to support the involvement
of traffic sources in an attack towards a target, while dealing with
the proper allocation of bandwidth during the attack via TE. The attacker's
classification of the target and decoys, as well as the attacking
bots should become more and more visible after a number of attack/re-route
interactions. We substantiate this model analytically in Section~\ref{sec:Analysis}.

\section{Analysis\label{sec:Analysis}}

Below we formulate the detection problem via relational algebra and
study the effects of the attacker-defender interplay. Then, the defender's
actions are analytically optimized.

\textbf{Prerequisites}. Assume a directed multigraph $G(\mathcal{N},\,\mathcal{L})$
comprising a set of nodes $\mathcal{N}$ and a set of directed links
$\mathcal{L}$. A single node is denoted as $n\in\mathcal{N}$ and
a single link as $l\in\mathcal{L}$. Each node $n$ logically hosts
a set of network-wide unique entities, $\mathcal{E}^{(n)}$. An entity
$e$ represents an IP address or an IP prefix, depending on the required
level of detection granularity. Each $\mathcal{E}^{(n)}$ set is connectable
to any other entity $e\in\mathcal{E}^{(n^{*})},\,n\ne n^{*}$; inter-entity
paths are based on the current set of routing tables $\mathcal{T}$
of $G$, i.e., the routing configuration.

The attack detection process will eventually relate a set of entities
(e.g., botnet IP addresses) to a set of nodes (e.g., decoy servers).
Therefore, we denote the associative relation, $r$, between two generic
sets $S_{1}$ and $S_{2}$ as:
\begin{equation}
r:\,S_{1}\stackrel{s}{\longrightarrow}S_{2}\label{eq:relationDef}
\end{equation}
where $s\in\mathbf{R}^{+}$ is the support of the relation. In general,
the $\longrightarrow$ notation denotes ordered collocations of $S_{1}$
and $S_{2}$ elements within a stream of $2$-tuple observations.
The support $s$ essentially counts these collocations within the
stream.

Let $\left\Vert S\right\Vert $ denote the cardinality of a set $S$.
We define the $L-specificity$ of relation (\ref{eq:relationDef})
as $\left\Vert S_{1}\right\Vert $, and its $R-specificity$ as $\left\Vert S_{2}\right\Vert $.
By definition, the following holds:
\begin{equation}
r_{1}\supseteq r_{2}\Leftrightarrow\left(S_{1}\to S_{2}\right)\supseteq\left(S_{3}\to S_{4}\right)\Rightarrow\left\{ \begin{array}{c}
S_{1}\supseteq S_{3}\\
S_{2}\supseteq S_{4}
\end{array}\right.
\end{equation}

\subsection{Problem Formulation via Associative Relations}

The analysis assumes a cycle of attacks followed by defense actions.
We study only the time moments when links have been flooded; these
moments are denoted as sequential time steps $t\in\mathbf{N}$. In
addition, the described process may \emph{optionally} assume the existence
of heuristics that classify link floods as malevolent, filtering out
natural causes like flash crowds \cite{thapngam2011discriminating}.

An attack at time step $t$ floods a set of links $\mathcal{L_{A}}(t)\subset\mathcal{L}$,
affecting the connectivity of a set of nodes $\mathcal{N_{A}}(t)\subset\mathcal{N}$.
Let $\mathcal{E_{A}}(t)$ denote the set of all \emph{suspicious}
entities, defined as the ones that are origins of traffic flows present
in $\mathcal{L_{A}}(t)$ which did not exist at time $t-1$. In other
words, $\mathcal{E_{A}}(t)$ contains the entities that have launched
new flows to be present in one or more flooded links at time step
$t$. Thus, for all time steps up to $t$, we form the following associative
relation:
\begin{equation}
r_{\mathcal{A}}(t):\,\underset{\forall t}{\cup}\mathcal{E_{A}}(t)\longrightarrow\underset{\forall t}{\cup}\mathcal{N_{A}}(t)\label{eq:DeltaDestin}
\end{equation}
The goal of the defender at time step $t$ is to deploy a new set
of routing tables effective until $t+1$, i.e., $\mathcal{T}(t+1)$,
such that:
\begin{equation}
min\left\{ s\right\} \,\forall r:\,e\stackrel{s}{\to}n,\,r\subset r_{\mathcal{A}}(t)\label{eq:formulation}
\end{equation}
The defender assumes that all entities $e\in\underset{\forall t}{\cup}\mathcal{E_{A}}(t)$
are benevolent, i.e., flows originating from an entity $e\in\mathcal{E_{A}}(t)$
will not change their destination to affect $n$ at time step $t+1$,
to the extent justified statistically by normal traffic patterns.
Thus he seeks to minimize the corresponding support to match this
assumption. In essence, the formulation of relation (\ref{eq:formulation})
describes the construction of new routing tables, $\mathcal{T}(t+1)$,
that disrelate entities from nodes affected by link-flooding events.
The new $\mathcal{T}(t+1)$ assumes that all past $e\to n$ relations
were coincidental and, therefore, their support should decrease with
high probability in the future. Relations that persist regardless
of this effort are treated as indications of an attack.

\textbf{Definition of Support}. At any given time step $t$, the defender
observes a set of $\mathcal{L_{A}}(t)$ flooded links. With no loss
of generality, let $\mathcal{L_{A}}(t)=\left\{ l_{i},\,i=1\ldots\left\Vert \mathcal{L_{A}}\right\Vert \right\} $.
Let $\mathcal{E}_{i}$ be the set of suspicious entities over link
$l_{i}$. We form the relations:
\begin{equation}
\left\{ r_{i}(t):\,\mathcal{E}_{i}\to l_{i}\right\} ,\,i=1\ldots\left\Vert \mathcal{L_{A}}\right\Vert \label{eq:EtoL}
\end{equation}
Moreover, let $\mathcal{N}_{i}$ denote the set of nodes which may
receive traffic via link $l_{i}$ according to $\mathcal{T}(t)$,
either as transit nodes or as flow destinations. We express these
relations as:
\begin{equation}
\left\{ r_{i}^{'}(t):\,l_{i}\to\mathcal{N}_{i}\right\} ,\,i=1\ldots\left\Vert \mathcal{L_{A}}\right\Vert \label{eq:LtoN}
\end{equation}
Relations (\ref{eq:EtoL}) and (\ref{eq:LtoN}) can be combined by
eliminating $l_{i}$:
\begin{equation}
\left\{ r_{i}^{''}(t):\,\mathcal{E}_{i}\to\mathcal{N}_{i}\right\} ,\,i=1\ldots\left\Vert \mathcal{L_{A}}\right\Vert \label{eq:EtoN}
\end{equation}

\begin{rem}
\label{rem:OverlappingRelations}The set $r_{i}^{''}(t)$ is used
for filtering out links from $\mathcal{L_{A}}(t)$ that carry the
same amount of information in terms of attack detection. An attack
may flood several links that are sequential on a path, due to link
capacity or link load variations. In these cases, the detection process
should consider the most general of the $r_{i}^{''}(t)$ relations
only, in order not to lose information that is important to the attack
detection process. That is, links that are ``shadowed'' by other
links on a path and do not offer new insight should be left out. Therefore,
the remainder of the analysis will assume that all links $l_{i}$
for which $\exists j:\,r_{j}^{''}(t)\supseteq r_{i}^{''}(t)$ have
been filtered out of $\mathcal{L_{A}}(t)$.
\end{rem}
For each distinct $e\in\bigcup_{\forall i}\mathcal{E}_{i}$, (\ref{eq:EtoL})
can be rewritten as:
\begin{equation}
e\stackrel{s(e)}{\longrightarrow}\mathcal{L}_{e}(t)=\left\{ l_{i}:\,e\in\mathcal{E}_{i}\right\} ,\,s(e)=\frac{\left\Vert \mathcal{L}_{e}(t)\right\Vert }{\left\Vert \mathcal{L_{A}}(t)\right\Vert }\label{eq:EtoFLOOD}
\end{equation}
In a similar fashion, for each distinct $n\in\bigcup_{\forall i}\mathcal{N}_{i}$,
(\ref{eq:LtoN}) yields:
\begin{equation}
n\stackrel{s(n)}{\longrightarrow}\mathcal{L}_{n}(t)=\left\{ l_{i}:\,n\in\mathcal{N}_{i}\right\} ,\,s(n)=\frac{\left\Vert \mathcal{L}_{n}(t)\right\Vert }{\left\Vert \mathcal{L_{A}}(t)\right\Vert }\label{eq:NtoFLOOD}
\end{equation}
Combining (\ref{eq:EtoFLOOD}) and (\ref{eq:NtoFLOOD}) produces the
specific relation:
\begin{equation}
e\stackrel{s(e,n)}{\longrightarrow}n,\,s(e,n)=\frac{{\scriptstyle \left\Vert \mathcal{L}_{e}(t)\cap\mathcal{L}_{n}(t)\right\Vert }}{{\scriptstyle \left\Vert \mathcal{L}_{e}(t)\cup\mathcal{L}_{n}(t)\right\Vert -\left\Vert \mathcal{L}_{e}(t)\cap\mathcal{L}_{n}(t)\right\Vert }}.\label{eq:EtoNspecific}
\end{equation}
Notice that $s(e)$ expresses the probability of $e$ acting as a
bot at time $t$, $s(n)$ is the probability of $n$ being an attack
target at time $t$, and $s(e,n)$ is the probability of $e$ attacking
$n$ at time $t$. The optimization criterion (\ref{eq:formulation})
thus seeks to minimize the \emph{total} support of relations (\ref{eq:EtoFLOOD}-\ref{eq:EtoNspecific})
\emph{over all time steps}, i.e., the probabilities $\Pi_{\forall t}s_{t}(e)$,
$\Pi_{\forall t}s_{t}(n)$ and $\Pi_{\forall t}s_{t}(e,n)$.

\subsection{Effects of Attacker's Actions on the Detection Process}

A smart attacker may consider that he may be tracked. Thus, he needs
to take measures to obfuscate his presence. The existence of a bot
entity may be obfuscated by:
\begin{enumerate}
\item Not participating to an attack at every step $t$.
\item Attacking targets beyond the intended ones.
\end{enumerate}
These approaches affect the $L$ and $R$ specificity of the relations
(\ref{eq:EtoNspecific}), as shown in the following Lemmas.
\begin{lem}
\label{lem:Lspec}Limited entity participation to an attack reduces
the L-specificity of a relation $e\stackrel{s(e,n)}{\longrightarrow}n$. \end{lem}
\begin{IEEEproof}
Assume an attack towards node $n$, consistently launched by a an
entity $e$ over time steps $0$ to $t$, leading to the relation
$e\stackrel{s(e,n)}{\longrightarrow}n$. Assume that the same attack
is now launched by a multiset of entities, $\mathcal{E}$, with $\left\Vert \mathcal{E}\right\Vert =t$,
where an entity $e\in\mathcal{E}$ is allowed to attack at one time
step only. Any cycle of size $\left\Vert \mathcal{E}\right\Vert $
can be used to deduce the exact order. The updated relation is now
$\mathcal{E}\stackrel{s(e,n)}{\longrightarrow}n$, yielding the same
support, but a decreased $L$-specificity of $\left\Vert \mathcal{E}\right\Vert >1$.\end{IEEEproof}
\begin{lem}
\label{lem:Rspec}Sporadically attacking targets beyond the intended
one reduces the R-specificity of a relation $e\stackrel{s(e,n)}{\longrightarrow}n$. \end{lem}
\begin{IEEEproof}
Similar to Lemma \ref{lem:Lspec}, using a $\mathcal{N}$ multiset.
\end{IEEEproof}
Attacks employing Lemmas \ref{lem:Lspec} and \ref{lem:Rspec} can
also be combined to yield relations with $L$ and $R$ specificity
that is too low to be of any practical use. For example, such relations
may produce only coarse indications of the form: ``A network receives
too much load originating from beyond its Internet-facing gateways''.
Notice that an attack that employs only Lemma \ref{lem:Lspec} will
still be $R$-specific via relation (\ref{eq:NtoFLOOD}), potentially
enabling TE-based defense measures despite the unspecificity of the
attack sources. Similarly, an attack that only uses Lemma \ref{lem:Rspec}
for obfuscation will still be $L$-specific via relation (\ref{eq:EtoFLOOD}),
enabling direct mitigation, e.g., via entity filtering or blacklisting.

However, obfuscating an attack via Lemmas \ref{lem:Lspec} and \ref{lem:Rspec}
also comes with a price, regardless of their usage combination. An
attacker must have an increased number of entities at his disposal,
in order to efficiently use Lemma \ref{lem:Lspec}. Likewise, attacking
extraneous targets (see Lemma \ref{lem:Rspec}) requires the use of
additional entities as well. Given that the bandwidth (and the number
of entities) required to flood a link is closely related to the TE
choices of the victim, we proceed to study the effects of classic
TE objectives to the specificity of relations (\ref{eq:EtoNspecific}).

\subsection{Effects of Classic TE Objectives on the Detection Process\label{sub:Effects-of-standard}}

A usual TE objective is to react to link congestion/flooding events,
redistributing the network load as equally as possible among the available
routes \cite{Pioro.2004}. This goal is most commonly expressed as
the minimization of the maximum link utilization throughout the network,
given the present traffic flow between each node pair \cite{Balon.2006b}.
The process is split into two stages:

\textbf{1) Optimal calculation of path load}. This stage receives
as inputs the current traffic matrix and a set of possible paths connecting
each pair of networked entities. It then produces the optimal aggregate
load that each route should carry. This problem can be formulated
as a Linear Program (LP) as follows. Let $M_{e,e'}$ be the current
traffic matrix containing the data rates between any two entities
$e$ and $e'$. Let the triplet $\left\langle e,e',k\right\rangle ,\ k=1\ldots K$
index each of the $K$ available paths that can connect the entities
$e,\ e'$ ($K$ can be a function of $e,\ e'$). Furthermore, let
$f_{\left\langle e,e',k\right\rangle }$ represent the fraction of
$M_{e,e'}$ over path $\left\langle e,e',k\right\rangle $. The objective
of stage $1$ is then achieved as follows \cite{Curtis.2011b}:
\begin{equation}
\begin{array}{cc}
\text{minimize:} & \mathcal{U}\\
\text{subject to:}\\
\forall_{e,e'}: & \sum_{\forall k}f_{\left\langle e,e',k\right\rangle }=1\\
{\scriptstyle \forall\left\langle e,e',k\right\rangle ,\forall l\in\left\langle e,e',k\right\rangle :} & \sum_{\forall e,e'}M_{e,e'}f_{\left\langle e,e',k\right\rangle }\le\mathcal{U}\cdot C_{l}
\end{array}\label{eq:LP}
\end{equation}
where $C_{l}$ is the nominal capacity of link $l$ and $\mathcal{U}\in\left[0,1\right]$
a helper variable. The first condition expresses the conservation
of the traffic load, while the second one ensures that the load of
each link is within its capacity constraint.

\textbf{2) Mapping of entity pairs to paths}. Given the current traffic
matrix $M_{e,f}$, this stage maps pairs of entities $\left\langle e,e'\right\rangle $
to paths $\left\langle e,e',k\right\rangle $ in order to match the
optimal $f_{\left\langle e,e',k\right\rangle }$ values produced in
stage $1$, after running the LP of formulation (\ref{eq:LP}).

It is worth noting that there exist single-stage TE approaches based
on metaheuristics (e.g., Genetic Algorithms), which attempt the same
optimization \cite{Ericsson.2002,Buriol.2005}. However, their performance
is generally inferior to the LP-based approach.

The following remarks can be made on the relation between the TE module
and the attack detection process. i) Since the TE is triggered on
link congestion events, it will also be the first to respond to link
flooding attacks as well. Therefore, the TE is a promising point for
introducing detection-oriented mechanisms. ii) It is mandatory that
the TE relieves the congested links, regardless of any actions pertaining
to the attack detection process. iii) The second stage of TE, i.e.,
the entity pair to path mapping, relates entities to nodes and can
potentially be tuned to aid detection, without changing the traffic
distribution derived in stage $1$. In light of these remarks, we
proceed to study the distinct effects of minimizing the maximum link
load (TE stage $1$) on the detection process, regardless of any stage
$2$ TE approach. The tuning of TE stage $2$ to detection purposes
follows in Section \ref{sub:Incorporation}, based on relation extraction.

\begin{figure}[!t]
\begin{centering}
\includegraphics[bb=0bp 0bp 490bp 240bp,clip,width=0.75\columnwidth,height=3cm]{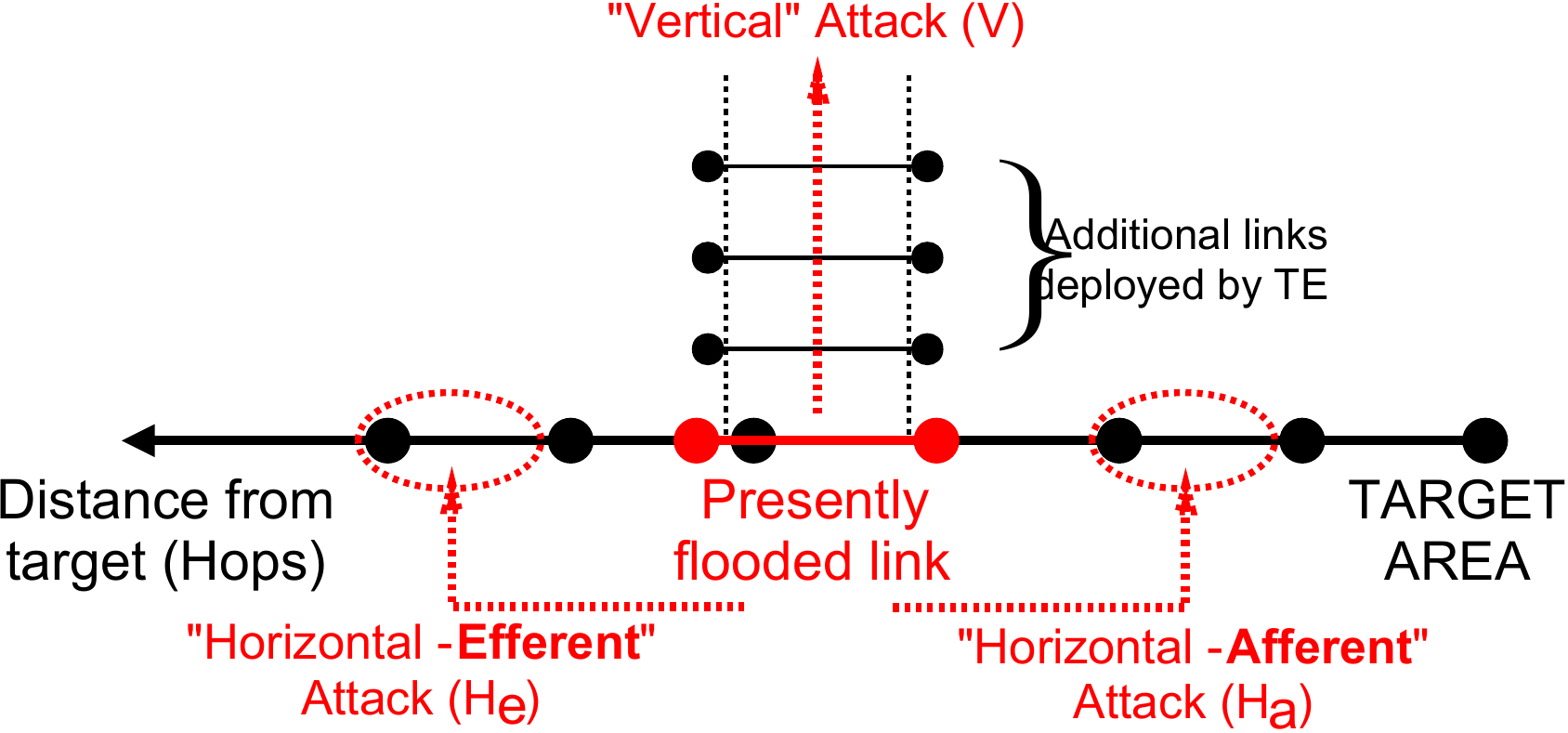}
\par\end{centering}

\protect\protect\caption{\label{fig:attackTypes}{\footnotesize{}Types of attack responses
to a link load-minimizing TE scheme. An attacker may move: i) ``vertically'',
flooding the additional links deployed via TE, or ii) ``horizontally'',
either nearing or distancing from the target area.}}
\end{figure}

Assume that an attack floods a link $l$ with capacity $C_{l}$ affecting
a target node, as shown in Fig. \ref{fig:attackTypes}. At this point,
let the aggregate attack traffic be $M\le C_{l}$. A link load-minimizing
TE scheme will reply by distributing the load of $l$ over additional
paths ``parallel'' to the congested. By definition of the min-max
link utilization objective, no new path has $100\%$ load after the
TE step (assuming a non-saturated network). Therefore, flooding at
least one of the parallel paths (including $l$) will require the
increase of the aggregate attack traffic to $M'\ge M$.

In response to the TE run, the attacker may choose to attack ``vertically'',
indeed flooding some or all of the deployed parallel links as required.
However, the bandwidth emanating from each of the attacker' entities
is upper bounded to avoid raising direct suspicions of flooding attempts
\cite{crossfire}. Therefore, it is valid to assume that the number
of entities participating to the attack, $n_{\mathcal{E}}$, increases
with their total coerced load, i.e.,:
\begin{equation}
M'\ge M\Rightarrow n_{\mathcal{E}}'\ge n_{\mathcal{E}}
\end{equation}
Assuming that the total number of entities available to the attacker
is upper bounded, increasing $n_{\mathcal{E}}$ means that each entity
must participate to attacks more frequently. Therefore, in accordance
with Lemma \ref{lem:Lspec}, we conclude:
\begin{prop}
\label{prop:Vertical-attack}Vertical attacks favor the $L$-specificity
of detected associative relations ($L^{+}$).
\end{prop}
The effect of vertical attacks to $R$-specificity may not be deduced
independently of the $2^{nd}$ TE stage. The newly deployed parallel
paths may simply implicate more nodes than those present at the right
side of an existing $\mathcal{E}\to\mathcal{N}$ relation, decreasing
$R$-specificity. However, other mappings may, e.g., divide $\mathcal{N}$
into node subsets that are routed via link-disjoint paths, increasing
$R$-specificity in the face of a vertical attack.

Alternatively, an attacker may choose to reply ``horizontally'',
towards the ``efferent'' direction ($H_{e}$), in an effort to deliberately
decrease the $R$-specificity of $\mathcal{E}\to\mathcal{N}$ relations.
Links further from the target area are likely to affect the routing
of more nodes, increasing $\left\Vert \mathcal{N}\right\Vert $. However,
such links will have greater nominal capacity for the same reason.
This hypothesis is further reinforced by the over-provisioned nature
of backbone links, capacity-wise. It is not uncommon for links to
have a nominal capacity several times greater than the traffic demands
of the served nodes \cite{zhang2004designing}. Thus, attacking links
towards $H_{e}$ is expected to require increased entity participation
in the general case. Therefore, similarly to Proposition \ref{prop:Vertical-attack}
we conclude the following:
\begin{prop}
\label{prop:He-attack}Horizontal-efferent attacks increase the $L$-specificity
and decrease the $R$-specificity of detected associative relations
($L^{+},\,R^{-}$).
\end{prop}
Working similarly, we deduce the duality of the ``Horizontal-afferent''
attacks, i.e., attacks to links closer to the target area:
\begin{prop}
\label{prop:Ha-attack}Horizontal-afferent attacks decrease the $L$-specificity
and increase the $R$-specificity of detected associative relations
($L^{-},\,R^{+}$). \end{prop}
\begin{rem}
\label{rem:notTooClose}We note that an attack must be horizontally
bounded, i.e., not moving indefinitely towards $H_{a}$ or $H_{e}$.
Flooding links too near to the target area may give away the attack
by minimizing the $R$-specificity of the detected relations. On the
other hand, attacks too far from the target may actually miss their
objective, affecting the connectivity of other nodes instead of the
intended target, or demand too much bandwidth.
\end{rem}
\begin{table}[t]
\centering{}\protect\protect\caption{\label{tab:Effects-of-attack}{\footnotesize{}Effects of attack responses
on the $L$ and $R$ specificity of a detected entity$\to$target
relation.}}

\centering{}%
\begin{tabular}{|c||c|c||c|c||c|}
\hline
\multicolumn{6}{|c|}{Attacker's response types}\tabularnewline
\hline
\multicolumn{2}{|c|}{$V$} & \multicolumn{2}{c|}{$H_{e}$} & \multicolumn{2}{c|}{$H_{a}$}\tabularnewline
\hline
\hline
\multicolumn{2}{|c|}{$L+$, $R\pm$} & \multicolumn{2}{c|}{$L+$, $R-$} & \multicolumn{2}{c|}{$L-$, $R+$}\tabularnewline
\hline
\end{tabular}
\end{table}

The effects of the attacker's responses are summarized in Table \ref{tab:Effects-of-attack}.
Notice that concurrent combinations of response types over a given
path are also possible. In this case, however, due to Remark \ref{rem:OverlappingRelations},
the associative relations $r$ observed after the attacker's response
will overlap as $r_{H_{e}}\supseteq r_{V}\supseteq r_{H_{a}}$, yielding
the specificity effects of the efferent-most attack. We observe that
each of the attacker's responses yields a gain in either $L$ or $R$-specificity,
enforcing at least one of the relations (\ref{eq:EtoFLOOD}), (\ref{eq:NtoFLOOD})
or (\ref{eq:EtoNspecific}). Therefore:
\begin{lem}
\label{lem:min-max-favors}The first stage of a min-max link utilization
TE objective facilitates the detection of attacks at a given time
step by increasing the $L$ or $R$-specificity of the observed entity-to-target
relations.
\end{lem}
Notice that Lemma \ref{lem:min-max-favors} does not at first preclude
attack \emph{strategies},\emph{ }i.e., series of attacker responses
that may yield a solid loss in specificity over a time horizon. For
instance, if $H_{e}$ and $H_{a}$ constantly yield $\left\langle L^{+1},R^{-1}\right\rangle $
and $\left\langle L^{-10},R^{+1}\right\rangle $ (using arbitrary
units), then continuous alternations between the two responses would
produce an unbounded loss in $L$-specificity with no effects on the
$R$-specificity. However, the feasibility of such a strategy depends
on the $2^{nd}$ TE stage, which quantifies the exact losses/gains
in $L$/$R$-specificity. We therefore proceed next to directly optimize
the $2^{nd}$ TE stage according to formulation (\ref{eq:formulation}),
favoring the detection process.

\subsection{Incorporation of Associative Relation Extraction to Classic TE Formulations
\label{sub:Incorporation}}

Let $r_{t}:\,e\stackrel{s}{\to}n$ be an observed relation with total
support $s$ at time step $t$. The goal of formulation (\ref{eq:formulation})
is to minimize $s$ at time $t+1$, assuming that $r_{t}$ represents
a false positive.

Let $dest(e_{f},t)$ return the destination of any \emph{flow} $e_{f}$
originating from entity $e$ at time $t$, noticing that $dest(e_{f},t)\ne n$
in general, by definition of the studied Crossfire attacks~\cite{crossfire}.
In addition, let $H_{t}$ represent the history of the destinations
of $e_{f}$:
\begin{equation}
H_{t}:\,\left\{ dest(e_{f},\tau),\,\tau\in\left[1,t\right]\right\}
\end{equation}
Given $H_{t}$, a false-positive flow may retain its destination (affecting
$n$ or not) at $t+1$ with probability $P_{ret}(e_{f})=P\left(\left.dest\left(e_{f},t+1\right)=dest\left(e_{f},t\right)\right|H_{t}\right)$,
or alter it to any other node $m\ne dest\left(e_{f},t\right)$ (including
null) with probability $P_{m}(e_{f})=P\left(\left.dest\left(e_{f},t+1\right)=m\right|H_{t}\right)$.
Let $P_{ret}^{a}(e_{f})$ and $P_{m}^{a}(e_{f})$ denote the probabilities
of a flow $e_{f}$ affecting $n$ again at time $t+1$ due to this
normal behavior justified by $H_{t}$. Minimizing the support $s$
is then tantamount to minimizing the total probability of re-enforcing
the relation $e\to n$ at $t+1$:
\begin{equation}
P_{t+1}^{a}(e)=\sum_{\forall e_{f}}\left(P_{ret}(e_{f})\cdot P_{ret}^{a}(e_{f})+\sum_{\mathclap{\forall m\ne dest\left(e,t\right)}}P_{m}(e_{f})\cdot P_{m}^{a}(e_{f})\right)\label{eq:pEffTotal}
\end{equation}
Firstly, we proceed to minimize the term $P_{ret}(e_{f})\cdot P_{ret}^{a}(e_{f})$.
Note that the term $P_{ret}(e_{f})$ is invariant to any defender's
actions, while $P_{ret}(e_{f})\ne0$ in the general case. We thus
proceed to minimize the term $P_{ret}^{a}(e_{f})$:
\begin{rem}
\label{rem:map1}$P_{ret}^{a}(e_{f})=0$ when the pairs $\left\langle e_{f},dest(e_{f},t)\right\rangle $
and $\left\langle origin,n\right\rangle $ are mapped to link-disjoint
paths in $\mathcal{T}(t+1)$.
\end{rem}
The identity of the $origin$ (Fig.~\ref{fig:crossfireOverfiew})
may not be known, especially during the first attack rounds. Therefore,
Remark~\ref{rem:map1} is practically implemented by mapping $\left\langle e_{f},dest(e_{f},t)\right\rangle $
to a path that is link-disjoint from $\left\langle s\left(l\right),n\right\rangle \,\forall flooded(l)$
at time $t$, $s\left(l\right)$ being the source node of $l$.

Secondly, we attempt to minimize the term $\underset{\forall m}{\sum}P_{m}(e_{f})\cdot P_{m}^{a}(e_{f})$
of equation (\ref{eq:pEffTotal}). Let $\mathcal{M}(t+1)$ be the
set of possible destinations $m$ of $e_{f}$, $m\ne dest(e_{f},t)$,
affecting $n$ in $\mathcal{T}(t+1)$ via one or more links (i.e.,
potential attack points). It holds that:
\begin{equation}
P_{m}(e_{f})\cdot P_{m}^{a}(e_{f})=\sum_{\mathclap{\forall m\in\mathcal{M}(t+1)}}P\left(\left.dest\left(e_{f},t+1\right)=m\right|H_{t}\right)\label{eq:PaltEff}
\end{equation}
In order to minimize equation (\ref{eq:PaltEff}), we employ virtual
links \cite{Koldehofe.2012} and proceed as follows:
\begin{lem}
\textup{\label{lem:map2}The configuration of the routing tables $\mathcal{T}\left(t+1\right)$
minimizes }$\underset{\forall m}{\sum}P_{m}(e_{f})\cdot P_{m}^{a}(e_{f})$\textup{
when the following conditions hold: i) It connects $origin$ to $n$
via a virtual link, ii) The virtual link comprises the smallest number
of physical links (i.e., shortest path), iii) The nodes along the
virtual link are the most improbable destinations of $e_{f}$ at time
$t+1$. }\end{lem}
\begin{IEEEproof}
Condition (i) ensures that the set $\mathcal{M}\left(t+1\right)$
of equation (\ref{eq:PaltEff}) contains only the intermediate nodes
encountered on the single, physical path represented by the virtual
link. Without virtual linking, any physical link on the path from
$origin$ to $n$ may serve any number of additional nodes, increasing
$\left\Vert \mathcal{M}\left(t+1\right)\right\Vert $. Condition (ii)
ensures that $\left\Vert \mathcal{M}\left(t+1\right)\right\Vert $
has the minimal value supported by the network. Condition (iii) then
minimizes $\underset{\forall m\in\mathcal{M}(t+1)}{\sum}P\left(\left.dest\left(e_{f},t+1\right)=m\right|H_{t}\right)$
by selecting the $\left\Vert \mathcal{M}\left(t+1\right)\right\Vert $
nodes with the lowest $P\left(\left.dest\left(e_{f},t+1\right)=m\right|H_{t}\right)$
values throughout the network. In order to circumvent the probably
unknown identity of the $origin$, we work as in Remark~\ref{rem:map1},
replacing $origin$ with $s(l)$ and repeating Lemma~\ref{lem:map2}
for every $flooded(l)$ at time $t$.
\end{IEEEproof}
Finally, the combination of Remark \ref{rem:map1} and Lemma \ref{lem:map2}
leads to the following Theorem.
\begin{thm}
\label{thm:optimalT}Let $r_{t}:\,e\stackrel{s}{\to}n$ be an associative
relation detected at time $t$. Let $e_{f}$ be a flow originating
from $e$ at time $t$. If the detection is falsely positive, the
routing $\mathcal{T}(t+1)$ that minimizes the support $s$ at time
$t+1$:

~i) Routes $\left\langle e_{f},dest(e_{f},t)\right\rangle $ and
$\left\langle origin,n\right\rangle $ via link-disjoint paths,~$\forall e_{f}$.
~ii) Routes $\left\langle origin,n\right\rangle $ via a virtual
link comprising the most improbable future destinations of $e$,~$\forall e_{f}$.
If all future destinations are equi-probable, $\left\langle origin,n\right\rangle $
is routed via the hop-wise shortest path between $origin$ and $n$.
\end{thm}
Qualitatively, Theorem \ref{thm:optimalT} facilitates detection by
forcing attacking entities to: i) constantly open new connections,
and ii) use improbable decoys in the process. Notice that it also
covers the total support of a relation up to $t+1$, which is expressed
as the probability $\mathbb{P}_{t+1}^{a}(e)=\prod_{\tau=1}^{t+1}P_{\tau}^{a}(e)$.
The theorem greedily reduces $\mathbb{P}_{t+1}^{a}(e)$ by minimizing
each of the $P_{\tau}^{a}(e)$ terms.

Regarding its incorporation to a min-max link utilization TE scheme,
the theorem can be implemented at the \emph{entity pair to path mapping}
phase (TE stage $2$). This phase maps each of the entity pairs to
one of the physical paths connecting them, keeping the total load
of each path near its optimal value. For each $e_{f}$, we first map
$\left\langle e_{f},dest(e_{f},t)\right\rangle $ to an arbitrary
path. Then, we proceed to list all available link-disjoint paths for
$\left\langle origin,n\right\rangle $, following Theorem \ref{thm:optimalT}.
If a pair of link-disjoint paths is not supported by the topology,
we select paths with joint links as close as possible to $origin$
or $n$, employing Remark \ref{rem:notTooClose}. Finally, we map
$\left\langle origin,n\right\rangle $ to the returned path with the
lowest $\underset{\forall m\in path}{\sum}P\left(\left.dest\left(e_{f},t+1\right)=m\right|H_{t}\right)$
value. Notice that flows with common source and destination nodes
are treated as an aggregated flow, via the same routing rules. Finally,
all non-suspicious entity pairs are then mapped to links arbitrarily.
The process has an average complexity of $O\left(\left\Vert \mathcal{R}_{t}\right\Vert \cdot K\cdot S\cdot F\right)$,
where $\mathcal{R}_{t}$ is the set of observed relations $r_{t}$,
$K$ is the average number of physical paths per node pair, $S$ is
the average number of links comprising a path, and $F$ is the average
number of flows originating from any entity.

Theorem \ref{thm:optimalT} also provides an insight on the topological
attributes that affect the vulnerability of a network to Crossfire-like
link-flooding attacks. Consider a full mesh topology, where all nodes
are equi-probable flow destinations. According to part (ii) of the
Theorem, the pair $\left\langle origin,n\right\rangle $ will always
be routed via the shortest paths. However, in a mesh topology these
paths always have a length of $1$ hop and, therefore, contain no
intermediate decoys. Thus, a Crossfire attack is not possible and
the network is invulnerable. In essence, the longer the paths offered
by a topology, the more the possible decoy nodes and the better the
probability of launching Crossfire successfully.

\subsection{On Reducing Complexity by Dissolving Infrequent Relations \label{sub:On-reducing-complexity}}

While Theorem \ref{thm:optimalT} can be implemented with linear complexity
w.r.t. the number of observed relations, certain special cases may
require additional attention. Firstly, an attacker may employ a very
large number of entities for an attack. Secondly, sizable networks
may yield too many probable attack targets. In addition, a single
entity may be used for attacking multiple targets at once, increasing
the number of observed relations further. Therefore, a defender may
need to reduce $\left\Vert \mathcal{R}_{t}\right\Vert $ by dissolving
some relations deterministically. The process is context-specific,
depending on the capabilities and requirements of the defender. For
instance, a top-$x$ approach can be employed, keeping only the $x$
relations with the greatest support. Another approach is to dissolve
relations that have not been observed within a given timeout period.
Another similar approach, advocated by well-known metaheuristics (e.g.,
ant colony optimization \cite{Luke.2013}) is to introduce a notion
of ``strength''. The strength of a relation (which is not related
to its support) is a number that increases every time the relation
is observed, but otherwise decreases as time elapses. If the strength
reaches a zero value, the relation is dissolved.

Regardless of the employed approach, the dissolution method must be
carefully selected, since it may work in favor of the attacker. For
instance, assume that an entity participates in each step of an attack
with probability $p$. Let $g(t)$ be the strength gathered by the
associated relations, which can be increased by $a>0$ and decreased
by $b>0$ at each step. Consider that $g$ is nullified for the first
time at $t+\tau$. Then, it must hold that:
\begin{equation}
g(t)+a\cdot p\cdot\tau-b\cdot(1-p)\cdot\tau=0\stackrel{g(t)\ge0}{\Rightarrow}\left[a\cdot p-b\cdot(1-p)\right]\cdot\tau<0
\end{equation}
Therefore, since $\tau>0$, we deduce that: $\,\,\,p<\nicefrac{b}{\left(a+b\right)}$.

In other words, if an attacker uses a $p$ value of less than $\nicefrac{b}{\left(a+b\right)}$,
then his bot relations and thus his attack will be stealthy.

Different dissolution approaches yield different behaviors. For instance,
if the defender follows a $\tau-$timeout approach, a relation will
be dissolved if not observed for $\tau$ time steps, i.e., with probability
$P=(1-p)^{\tau}$. Therefore, an attacker may choose a $p$ value
which ensures that $P$ is below a threshold. However, the attacker
should have the necessary number of entities $\left\Vert \mathcal{E}\right\Vert $
at his disposal to take advantage of dissolution in any case, taking
into account that his attacks should be successful with just $\left\Vert \mathcal{E}\right\Vert \cdot p$
entities. On the other hand, the defender should be aware of the trade-off
resulting from his relation dissolution approach, and carefully balance
computational complexity and detection potential.

\section{Algorithmic formulation\label{sec:Algorithmic-formulation}}

\begin{algorithm}[t]
\protect\protect\caption{\label{alg:ariel}{\footnotesize{}The proposed ARIEL scheme.}}

{\small{}\hrule \KwIn{The flooded links $\mathcal{L_{A}}$ at time
$t$; The entities $\mathcal{E}_{l}$ with new connections present
in each $l\in\mathcal{L}$; The nodes $\mathcal{N}_{l}$ affected
by each $l\in\mathcal{L}$; The traffic matrix $M$.} \KwOut{The
relations (\ref{eq:EtoL}), (\ref{eq:LtoN}), (\ref{eq:EtoN}) at
time $t$; the new routing configuration at time $t+1$, i.e., $\mathcal{T}(t+1)$.}
\hrule \tcc{Initialization.} Create persistent database tables
$R_{EL}$, $R_{LN}$\; \hrule \tcc{Remove shadowed links.} \ForEach{$l\in\mathcal{L_{A}}$}{
\If{$\exists l^{*}\in\mathcal{L_{A}}:\,\mathcal{E}_{l^{*}}\to\mathcal{N}_{l^{*}}\supseteq\mathcal{E}_{l}\to\mathcal{N}_{l}$}{
$\mathcal{L_{A}}\dashleftarrow\mathcal{L_{A}}-l$\; } } \tcc{Update
Database.} \ForEach{$l\in\mathcal{L_{A}}$}{ $\ensuremath{R_{EL}}\dashleftarrow\ensuremath{R_{EL}}\bigcup\left\{ \left\langle \mathbb{\epsilon}=e,\mathbb{\lambda}=l,\mathbb{\tau}=t\right\rangle ,\,\forall e\in\mathcal{E}_{l}\right\} $\;
$\ensuremath{R_{LN}}\dashleftarrow\ensuremath{R_{LN}}\bigcup\left\{ \left\langle \lambda=l,\nu=n,\tau=t\right\rangle ,\,\forall n\in N_{l}\right\} $\;
} \tcc{Produce Relations.} $\forall e\in R_{EL}:\,e\stackrel{s}{\to}\star\ensuremath{,\,s=\left\Vert \sigma_{\epsilon=e}R_{EL}\right\Vert }$\;
$\forall n\in R_{LN}:\,\star\stackrel{s}{\to}n\ensuremath{,\,s=\left\Vert \sigma_{\nu=n}R_{LN}\right\Vert }$\;
$\forall\left\langle e,n\right\rangle \in\left(\mathbf{R}:\,\ensuremath{R_{EL}}\stackrel{\lambda}{\bowtie}R_{LN}\right):\,e\stackrel{s}{\to}n\ensuremath{,\,s=\underset{\mathclap{\left\langle \epsilon,\nu\right\rangle =\left\langle e,n\right\rangle }}{\left\Vert \sigma_{\,\,\,\,\,\,\,\,\,\,\,}\mathbf{R}\right\Vert }}$\;
\tcc{Dissolve Relations by TimeOut.} $\ensuremath{R_{EL}}\dashleftarrow\ensuremath{R_{EL}}-\sigma_{\tau<t-TimeOut}R_{EL}$\;
$\ensuremath{R_{LN}}\dashleftarrow\ensuremath{R_{LN}}-\sigma_{\tau<t-TimeOut}R_{LN}$\;
\tcc{Perform TE.} Calculate optimal load fractions $f$ via formulation
(\ref{eq:LP})\; Produce $\mathcal{T}(t+1)$ via Theorem \ref{thm:optimalT}\;
\hrule }{\small \par}
\end{algorithm}

The analytical findings of Section \ref{sec:Analysis} lead to the
formulation of Algorithm \ref{alg:ariel} (Associative RelatIon Extraction
aLgorithm - ARIEL). The formulation allows for a straightforward implementation
based on a relational database, exploiting the performance, stability
and scalability benefits of this mature technology (e.g., SQL \cite{SQLbook}).
Therefore, we employ the additional notation of $\sigma_{c}\mathbf{R}$
to express the selection of rows of a table $\mathbf{R}$ yielding
true to the binary predicate condition $c$. In addition, $\mathbf{R}_{1}\stackrel{col}{\bowtie}\mathbf{R}_{2}$
will denote the natural join of tables $\mathbf{R}_{1}$ and $\mathbf{R}_{2}$
on column $col$. ARIEL requires two persistent database tables, $R_{EL}$
and $R_{LN}$ with the columns detailed in lines $8-9$. The tables
are created at the initialization phase. Subsequent runs of ARIEL
utilize the same table instances as detailed next.

ARIEL is executed on a link-flooding event, after the execution of
any extra heuristics for flow classification (e.g., \cite{thapngam2011discriminating}).
The set containing the flooded links at the time of execution $t$
is denoted as $\mathcal{L_{A}}$. The entities present in each $l\in\mathcal{L_{A}}$
are denoted as $\mathcal{E}_{l}$, while $\ensuremath{\mathcal{N}_{l}}$
contains the nodes whose traffic is routed via $l$. The defender
may utilize common network logs to deduce which connections did not
exist at $t-1$ \cite{Chowdhury.2014}. ARIEL performs the task of
attack detection at lines $2-15$ and min-max link utilization TE
at lines $16-17$. Firstly, links that will lead to extraneous relations
are filtered out of $\mathcal{L_{A}}$, in accordance with Remark
\ref{rem:OverlappingRelations} (lines $2-6$). The process can be
completed with an average of $O\left(\left\Vert \mathcal{L_{A}}\right\Vert \cdot log\left\Vert \mathcal{L_{A}}\right\Vert \cdot E_{l}\left[\left\Vert \mathcal{N}_{l}\right\Vert +\left\Vert \mathcal{E}_{l}\right\Vert \right]\right)$
calculations, given that the process can be treated as a partial sorting
of the elements of $\mathcal{L_{A}}$ by the $\supseteq$ operator.
Each comparison operation then requires $O\left(E_{l}\left[\left\Vert \mathcal{N}_{l}\right\Vert +\left\Vert \mathcal{E}_{l}\right\Vert \right]\right)$
computations, $E_{l}\left[*\right]$ denoting the average of $*$
with regard to variable $l$. At steps $7-10$ ARIEL populates the
tables $R_{EL}$ and $R_{LN}$ by processing $\mathcal{E}_{l},\,\ensuremath{\mathcal{N}_{l}}$
as required by definition (\ref{eq:DeltaDestin}), while also adding
timestamp information. The complexity is $O\left(\left\Vert \mathcal{L_{A}}\right\Vert \cdot E_{l}\left[\left\Vert \mathcal{N}_{l}\right\Vert +\left\Vert \mathcal{E}_{l}\right\Vert \right]\right)$.

The detected associative relations are produced in lines $11-13$.
As explained in Section \ref{sec:Analysis}, an attacker may attempt
to obfuscate the attacking entities or the attack target. At line
$11$, ARIEL attempts to detect all suspicious entities, regardless
of target ($e\to\star$), in case the latter is obfuscated. For each
distinct entity $e$, the support is proportional to the counting
of all $R_{EL}$ entries containing $e$. A similar approach is followed
at line $12$ to detect possible attack targets, regardless of the
attacker's identity. Finally, the relations $e\to n$ are derived
from the natural join of $R_{EL}$ and $R_{LN}$ over the links $\lambda$,
at line $13$. Notice that the support of the produced relations is
\emph{not normalized} in $\left[0,1\right]$ as stated in equations
(\ref{eq:EtoFLOOD}-\ref{eq:EtoNspecific}). The normalization is
skipped since it is trivial and reduces the clarity of the presentation
of ARIEL. The complexity is $O\left(\left\Vert R_{EL}\right\Vert +\left\Vert R_{LN}\right\Vert \right)$,
assuming that the database uses a hash approach to implement the join~\cite{SQLbook}.

The algorithm then proceeds to dissolve ``old'' relations at lines
$14-15$ by removing the corresponding entries from the tables $R_{EL}$
and $R_{LN}$. The timeout approach is used as an example. Alternative
dissolution approaches can be freely used at this point. The complexity
is $O\left(\left\Vert R_{EL}\right\Vert +\left\Vert R_{LN}\right\Vert \right)$.

The TE task is accomplished at lines $16-17$. The complexity of the
optimal link load calculation (line $16$) depends on the employed
LP solver \cite{megiddo1986complexity}. Finally, the detection-oriented
mapping of entity pairs to paths follows Theorem \ref{thm:optimalT},
whose process and complexity has been described in Section~\ref{sub:Incorporation}.

\section{Simulations\label{sec:Simulations}}

We next perform simulations to study the effects of i) the TE phases
(load spreading, entity mapping), and ii) the topological attributes
on the specificity of relations extracted by ARIEL. The simulator,
which we plan to release as a free open-source application, is implemented
on the AnyLogic platform \cite{XJTechnologies.2015}.

\begin{figure}[t]
\begin{centering}
\includegraphics[width=1.05\columnwidth]{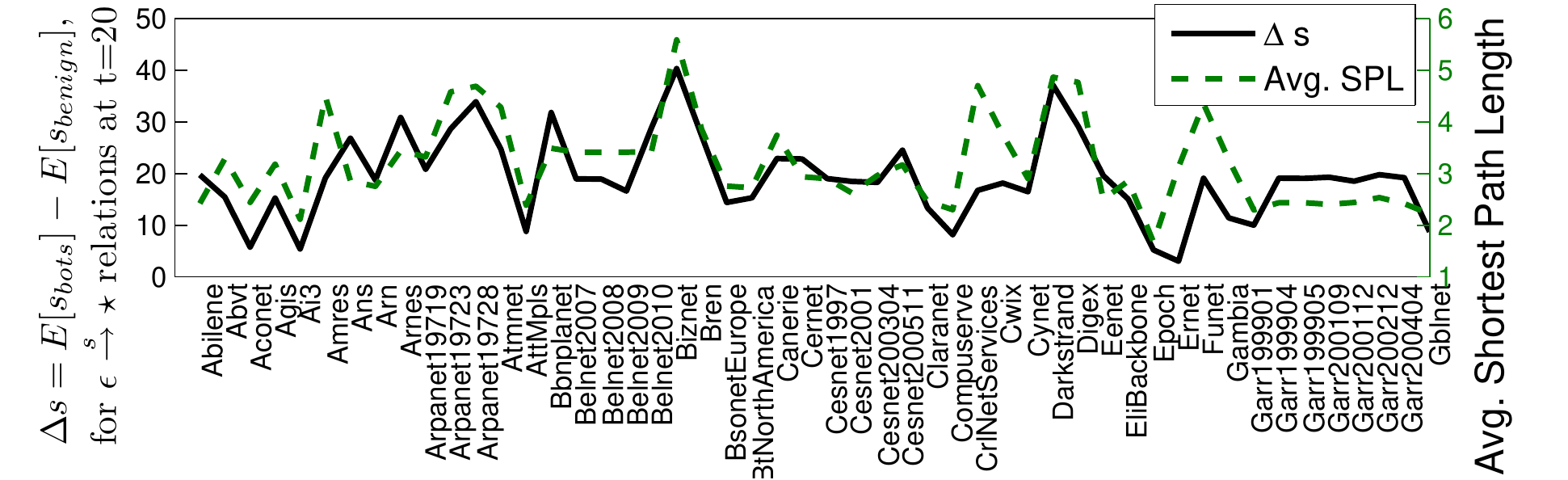}
\par\end{centering}

\protect\protect\caption{\label{fig:topo}{\footnotesize{}Effects of different topologies on
the specificity of the $\epsilon\stackrel{s}{\to}\star$ relations.
The effect is strongly correlated to the average-shortest-path-length
(avg. SPL) topology metric.}}
\end{figure}

\textbf{Setup.} The simulations assume one synthetic and $50$ real
topologies (listed in the x-axis of Fig. \ref{fig:topo}, derived
from the Internet Topology Zoo \cite{topologyzoo}). The synthetic
topology comprises $25$ nodes arranged in a $5\times5$ square grid.
In every topology, all links are set to a capacity of $10GBps$. Furthermore,
two alternative, link-disjoint paths are considered for each node-pair.
In addition, each topology hosts a flat number of $10,000$ benign
entities and $10,000$ bots, equally distributed to all nodes. This
selection corresponds to a botnet of considerable size, i.e., the
number of bots is equal to the total number of network users. Each
entity (benign or bot) is allowed to have up to $5$ connections opened
at any given time. The origin and the attack target are selected as
the most distant node pair (considering their hop-wise shortest path)
in each topology. Time $t$ is slotted, advancing at steps of $1$
on link-flooding events (as in Section \ref{sec:Analysis}), up to
$t=20$. The attacker operates as described in Section \ref{sec:The-Attack-Model}.
The bandwidth of each connection, $flow_{bw}$, is flat for bots or
benign entities. Given a simulation configuration, $flow_{bw}$ needs
to be calibrated to enable Crossfire. Too low $flow_{bw}$ is insufficient
for link-flooding, while too high means that the whole network is
flooded. We choose the lowest $flow_{bw}$ that enables the attack
and provide this value at each Figure.

For the sake of experimentation, we define the input parameters $reuse\_ratio$
and $rehome\_ratio$, pertaining to bot and benign \emph{connections}.
A $reuse\_ratio=10$ means that $10\%$ of the active bot connections
at time $t$, will re-adapt their destinations to participate in the
attack at time $t+1$. The remaining $90\%$ will remain unaltered.
Similarly, the $rehome\_ratio$ defines how many benign connections
will change their destination to a (uniformly) random node at time
$t+1$. Finally, the metric $\Delta s$ in introduced to quantify
the specificity of detected relations. $\Delta s$ is defined as the
average (not normalized) support of relations involving bots/target
nodes, minus the average support of relations involving benign entities/any
other node. A higher $\Delta s$ value means that bots/target nodes
stand out more and, therefore, can be detected more efficiently. Finally,
all runs are repeated for $95\%$ confidence in the results.

\begin{figure}[t]
\begin{centering}
\subfloat[\label{fig:botsAmassed}{\footnotesize{}Effects of TE phase 1 (load
balancing) on the bot availability. }]{\begin{centering}
\includegraphics[width=0.71\columnwidth]{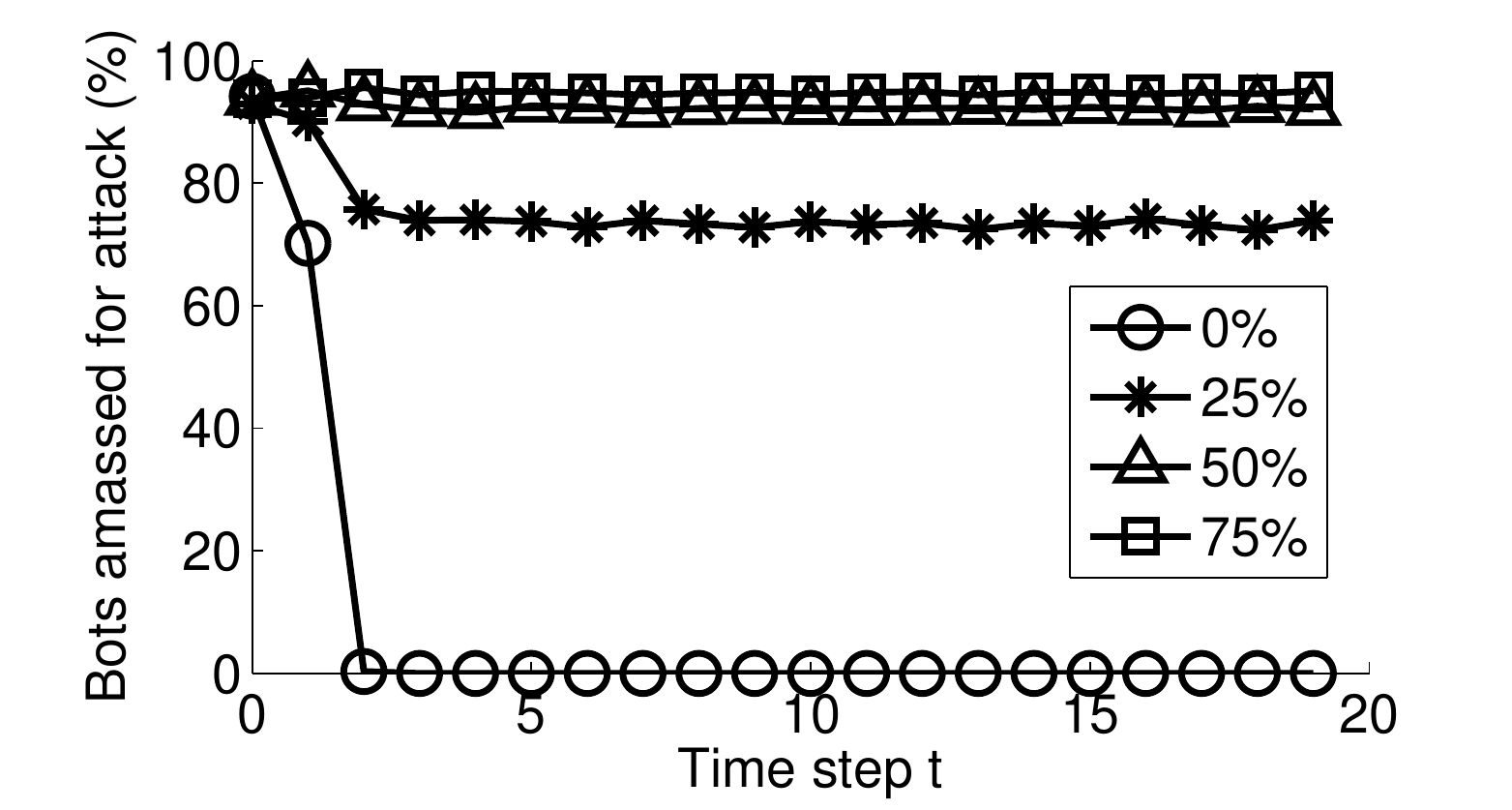}
\par\end{centering}

}
\par\end{centering}

\begin{centering}
\subfloat[\label{fig:attackSuccess}{\footnotesize{}Effects of TE phase 2 (flow
mapping to paths) on the attack success ratio.}{\small{} }]{\begin{centering}
\includegraphics[width=0.71\columnwidth]{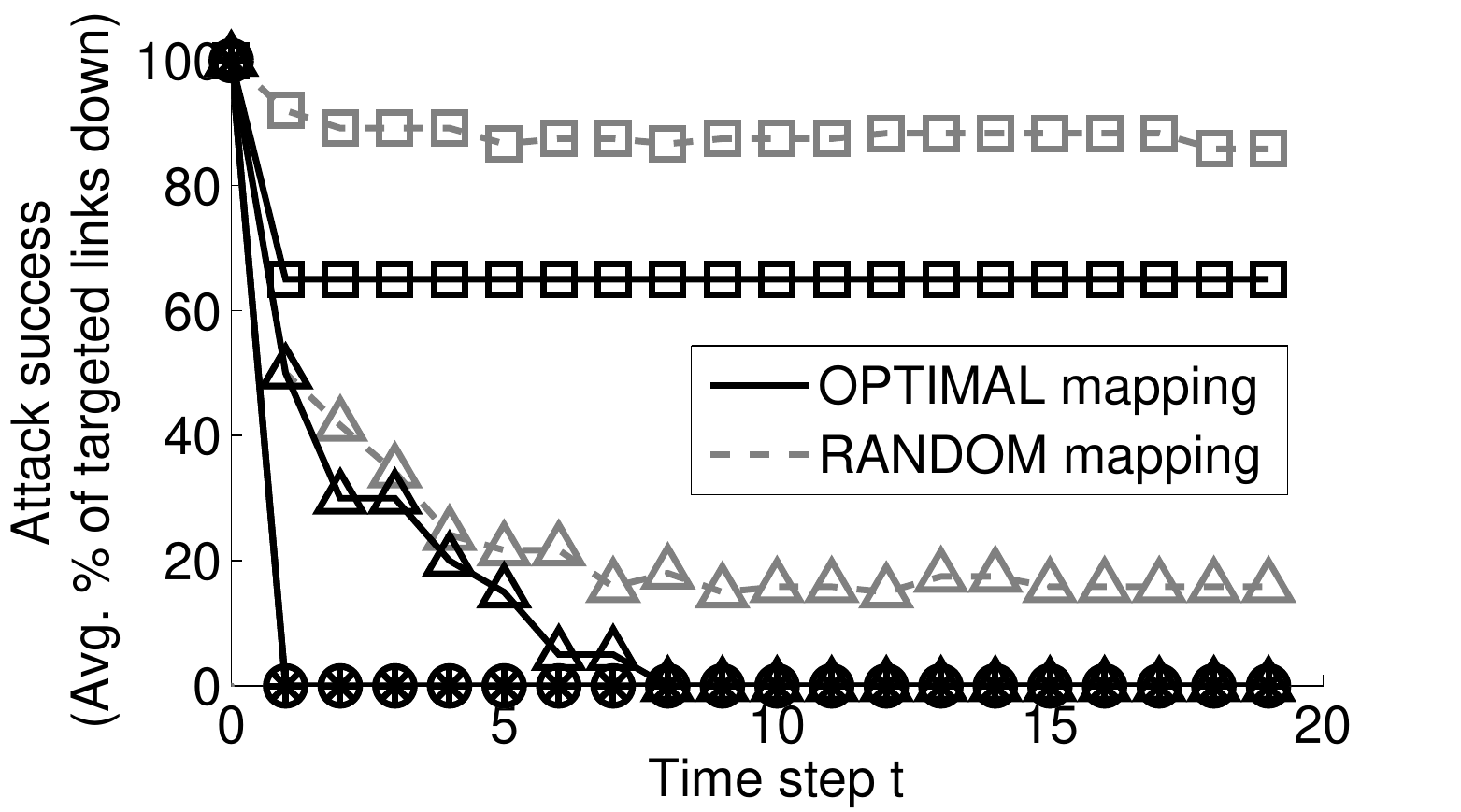}
\par\end{centering}

}
\par\end{centering}

\centering{}\protect\protect\caption{\label{fig:BotsSuccess}{\footnotesize{}The effects of two TE phases
on the number of bots drawn to an attack and its success ratio. The
$reuse\_ratio$ is varied in $0-75\%$. (Calibrated $flow_{bw}=400KBps$,
$rehome\_ratio=10\%$).}}
\end{figure}

\textbf{Results.} Figure \ref{fig:BotsSuccess} studies the general
effects of TE on the efficiency of Crossfire attacks. Using the synthetic
topology, we start with a bot connection $reuse\_ratio=0\%$. In just
two time steps, the attacker has run out of available bots (Fig. \ref{fig:botsAmassed})
and is unable to flood the targeted links (Fig. \ref{fig:attackSuccess}).
A $reuse\_ratio$ of $25\%$ increases the number of available bots
to $\approx75\%$, but their total connections are still insufficient
for a successful attack. Thus, the attacker is forced to a $reuse\_ratio=50\%$
which i) makes all bots visible (nearly $100\%$ participation, Fig.
\ref{fig:botsAmassed}), while ii) achieving marginally successful
attacks. It is at a $reuse\_ratio=75\%$ when the attacks become consistently
successful, should the defender use a random flow mapping at the $2^{nd}$
TE phase. The proposed optimal mapping makes even the $reuse\_ratio=75\%$
insufficient, forcing the attacker to: i) use all his bots, and ii)
use them almost exclusively for attacks, accentuating their detection.
\begin{figure}[t]
\begin{centering}
\includegraphics[width=0.71\columnwidth]{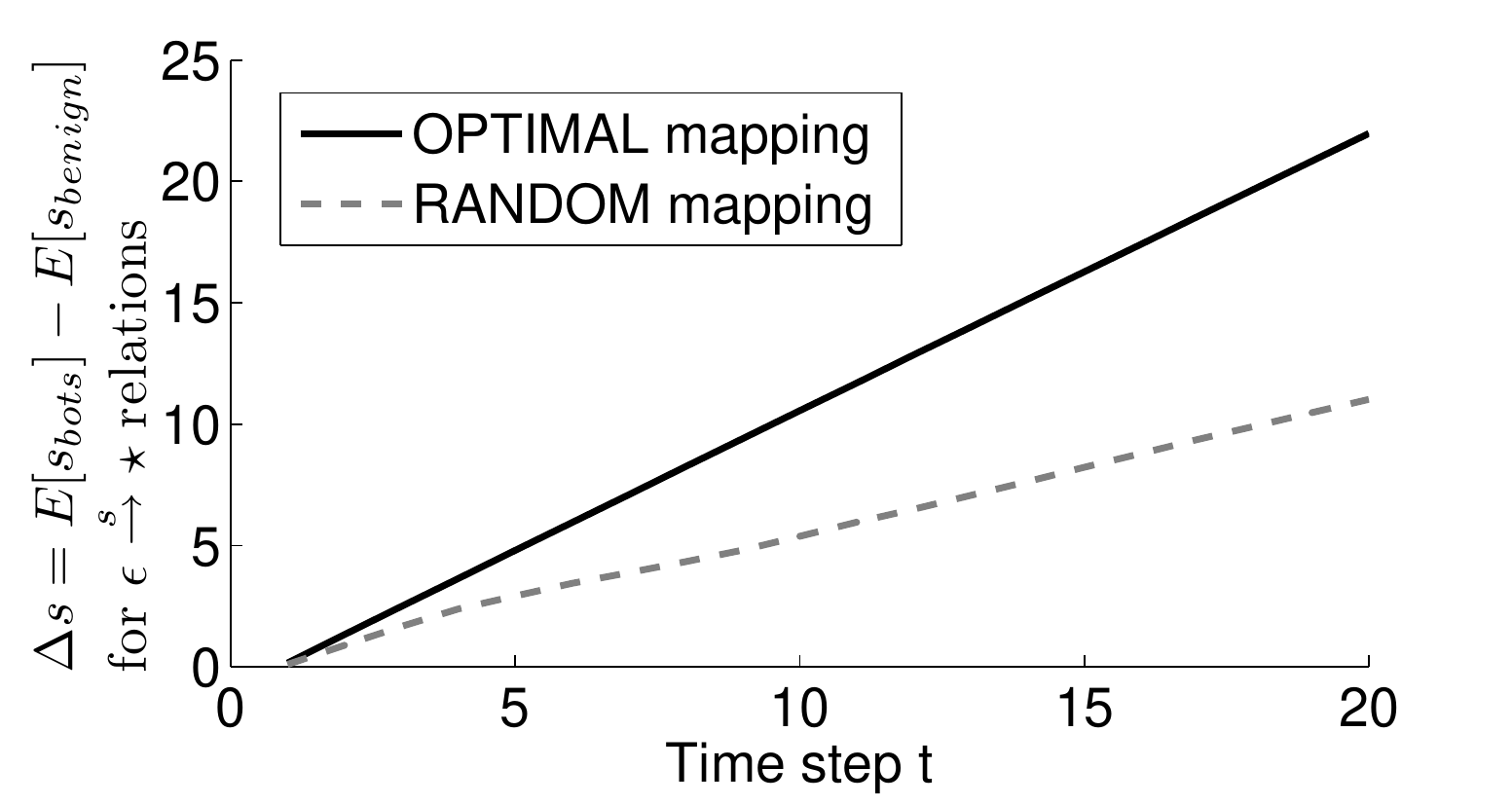}
\par\end{centering}

\protect\protect\caption{\label{fig:OPTIMALmapping}{\footnotesize{}Effects of optimal/random
mapping on the specificity of the $\epsilon\stackrel{s}{\to}\star$
relations. (Calibrated $flow_{bw}=400KBps$, $rehome\_ratio=10\%$).}}
\end{figure}

This becomes evident in Fig. \ref{fig:OPTIMALmapping}, where $\Delta s$
is doubled when the optimal mapping is used. We note that these findings
are aligned to the theoretical hypotheses and conclusions of Sections
\ref{sub:Effects-of-standard} and \ref{sub:Incorporation} on the
effects of TE on the detection process. Specifically, a load-balancing
TE is shown to naturally force an attacker to use more bots to remain
effective. In addition, the optimal mapping speeds up the detection.
Notice that the random mapping yields a strictly increasing specificity
as well, albeit at a slower rate.
\begin{figure}[t]
\begin{centering}
\includegraphics[width=0.71\columnwidth]{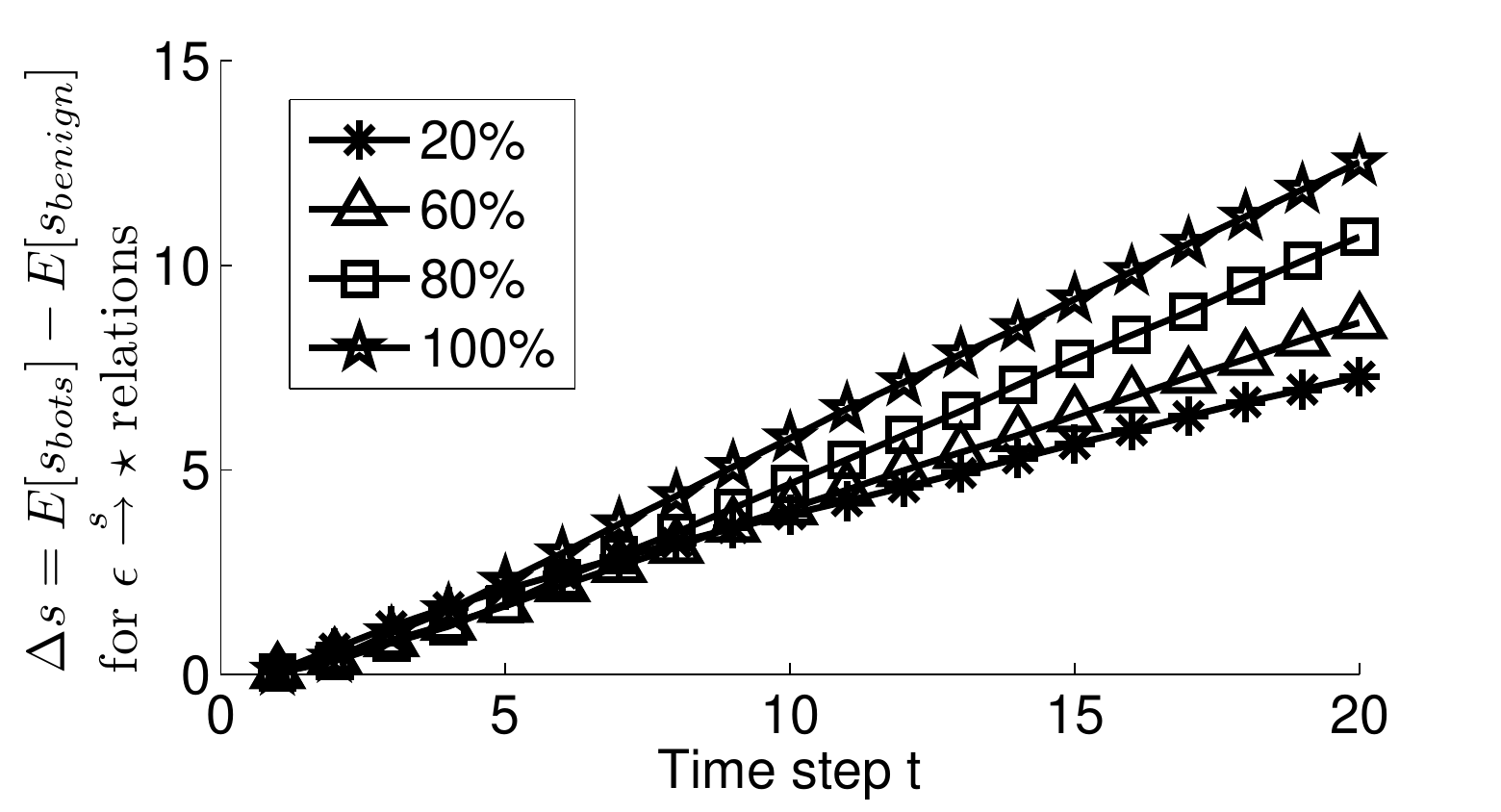}
\par\end{centering}

\protect\protect\caption{\label{fig:EtoANYspecificityEffects}{\footnotesize{}Effects of probabilistic
bot participation to an attack on the specificity of the $\epsilon\stackrel{s}{\to}\star$
relations. The $reuse\_ratio$ ratio is varied between $20-100\%$.
(Calibrated $flow_{bw}=1400KBps$, $rehome\_ratio=10\%$).}}
\end{figure}

\begin{figure}[t]
\begin{centering}
\includegraphics[width=0.71\columnwidth]{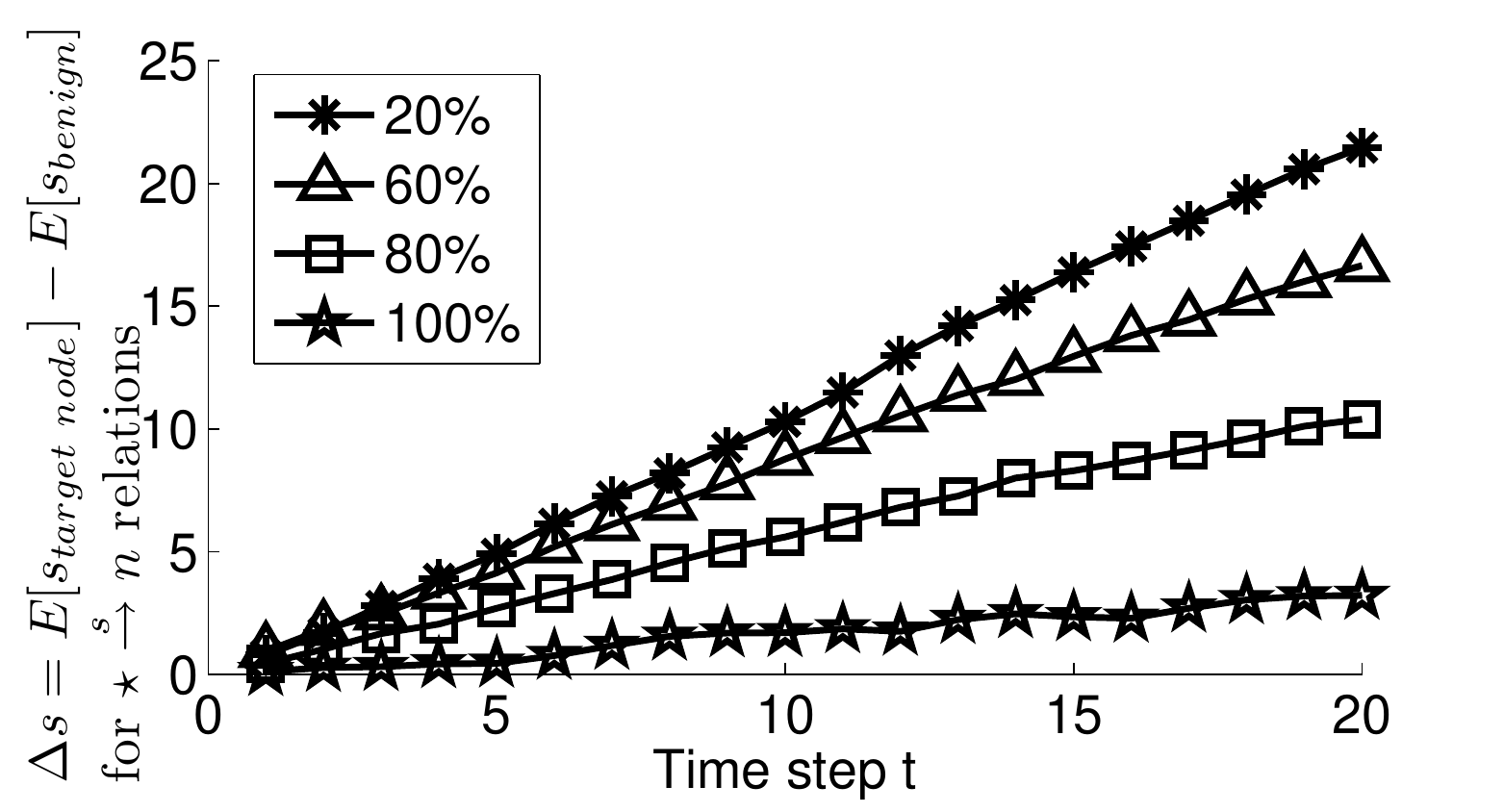}
\par\end{centering}

\protect\protect\caption{\label{fig:ANYtoNspecificityEffects}{\footnotesize{}Effects of attacking
random nodes (apart from the target) on the specificity of the $\star\stackrel{s}{\to}n$
relations. The $rehome\_ratio$ is varied between $20-100\%$. (Calibrated
$flow_{bw}=3500KBps$, $reuse\_ratio=100\%$).}}
\end{figure}

We proceed to study specifically the effectiveness of an attacker's
attempt to obfuscate his bots and his targets, corresponding to Lemmas
\ref{lem:Lspec} and \ref{lem:Rspec}. The $flow_{bw}$ is deliberately
increased to yield a very congested network, in order to make attacks
possible with less bots. Then, in Fig. \ref{fig:EtoANYspecificityEffects},
the $reuse\_ratio$ is varied from a high to a low value. As lower
values are used, the bots participate to the attacks less frequently,
which reduces the specificity of $\epsilon\to\star$ relations as
expected by Lemma \ref{lem:Lspec}. At this point we also note that
timing-out relations as described in Section \ref{sub:On-reducing-complexity}
has the exact same effect.

In Fig. \ref{fig:ANYtoNspecificityEffects} we set the $flow_{bw}$
to a value where the re-homing of benign connections causes link-flooding
events by itself. The higher the $rehome\_ratio$, the more the naturally
flooded links. In this manner, the attacker is expected to obfuscate
his targeted nodes, as stated by Lemma \ref{fig:ANYtoNspecificityEffects}.
The results validate the theoretical claim, and the specificity of
$\star\to n$ relations reduces when the $rehome\_ratio$ increases.

Finally, we proceed to test ARIEL in real topologies in Fig. \ref{fig:topo}.
The topologies are random, selected alphabetically by name in the
Topology Zoo database. Due to space restrictions, we illustrate the
achieved specificity of the $\epsilon\to\star$ relations, given that
the detection of the bots may be the first priority for the mitigation
of the attack. ARIEL yields positive $\Delta s$ in all cases, albeit
with varying end-value at $t=20$. To better understand the causes
of this behavior, we tested the correlation between $\Delta s(t=20)$
and several topology metrics (centrality, average number of neighbors,
average shortest path length-AvgSPL, number of nodes, diameter, clustering
coefficient, network density and network heterogeneity \cite{Killcoyne.2009}).
The AvgSPL metric exhibited the strongest covariance with $\Delta s$,
as shown in Fig. \ref{fig:topo}. Specifically, the two plots yield
a Pearson correlation coefficient of $0.7$, with $P=10^{-8}$ \cite{Benesty.2009}.
This outcome is aligned to Theorem \ref{thm:optimalT}, showing that
topologies that offer longer paths between nodes are more vulnerable
to attacks. However, the higher the vulnerability (higher AvgSPL),
the better the detection result. We note though that this correlation
is statistical and outliers exist (e.g., the ``CrlNetServices''
topology in Fig. \ref{fig:topo}). Nonetheless, the AvgSPL constitutes
a good metric for an initial estimation of the vulnerability of the
network, based on its topology.

\textbf{Future work}. We plan the following extensions. i) The formulation
of the attacker's responses (Table \ref{tab:Effects-of-attack}) paves
the way for a game-theoretic approach, where the goal is to derive
optimal attack and defense strategies. ii) Novel topological metrics
should be defined to quantify the vulnerability of a network deterministically.
iii) Noticing that the plots of Fig. \ref{fig:botsAmassed}-\ref{fig:ANYtoNspecificityEffects}
are very well-formed (e.g., linear), we target the derivation of the
exact formulas that govern the involved metrics.\vspace{-5bp}

\section{Related Work\label{sec:Related-Work}}

Studer et al. introduce the Coremelt attack \cite{Coremelt}, where
swarms of attack bots send traffic between each other in order to
cause significant congestion within core network links, as collateral
damage. The Crossfire attack~\cite{crossfire}uses bots as sources
and decoy servers as destinations, but falls within the same class
of attacks. Research around such attacks has focused mostly on the
system's side for detection. For example, Xue et al. propose the \emph{LinkScope}
system for detecting malicious link floods and for locating the target
link or area whenever possible \cite{xue2014towards}. Their system
uses end-to-end and hop-by-hop network measurement techniques to detect
abrupt degradation of performance. The survey of Bhuyan et al. \cite{bhuyan2013detecting}
gives an overview of the methods and tools used for detecting DDoS
attacks; these range from statistical methods to machine-learning
heuristic approaches. Zargar et al. \cite{zargar2013survey} further
present a comprehensive classification of various defense mechanisms
against DDoS flooding attacks. In contrast to these studies, we propose
a novel model and analysis of a joint detection and mitigation approach.

Several related DoS attack types have been studied in light of the
SDN paradigm shift as well. Braga et al. capitalize on controller
features for traffic analysis using Self Organizing Maps (SOMs) to
classify flows and enable DDoS attack detection caused by \emph{heavy
hitters} \cite{braga2010lightweight}. Ashraf et al. provide a general
survey of machine learning approaches for mitigating DDoS attacks
in SDN environments \cite{ashraf2014handling}. Lim et al. propose
a SDN-based scheme to block botnet-based DDoS attacks that do not
exhibit detectable statistical anomalies \cite{lim2014sdn}. The recent
work of Lee et al. (CoDef) \cite{codef} is a first approach towards
defeating new link-flooding attacks such as Coremelt and Crossfire.
The authors propose a cooperative TE-based detection method for identifying
low-rate attack traffic. The traffic sources and targets need to communicate
directly via an extra protocol. Malicious traffic sources are identified
by not complying with the instructed re-routing requests. Finally,
Gkounis \cite{gkounis2014thesis} studies the practical challenges
of using SDN to implement a joint detection and mitigation scheme.

\section{Conclusion\label{sec:Conclusions}}

This work introduced a novel framework for studying distributed link-flooding
attacks. The goal of the framework is to facilitate the detection
of susceptible bots and targeted network areas. This objective was
formulated in terms of relational algebra and was seamlessly incorporated
to standard TE modules. The analysis provided insights on optimizing
the detection process, isolating the impact of the attacker and facilitating
proper defender's reactions to the detection. Moreover, it shed light
on the topological attributes that significantly influence the vulnerability
of a network, attack-wise. The analytical insights were validated
via extensive simulations on a variety of real and synthetic topologies.

 \bibliographystyle{IEEEtran}

\end{document}